\documentclass[]{emulateapj}
\usepackage{natbib}
\usepackage{apjfonts}
\usepackage{epsfig}

\newcommand{\MBH}{M_{\mathrm{BH}}}
\newcommand{\Mbulge}{M_{\mathrm{bulge}}}
\newcommand{\msigma}{\MBH\mathrm{-}\sigma}
\newcommand{\mbulge}{\MBH\mathrm{-}\Mbulge}
\newcommand{\vir}{\mathrm{vir}}
\newcommand{\Mvir}{M_{\mathrm{vir}}}
\newcommand{\Vvir}{V_{\mathrm{vir}}}
\newcommand{\Rvir}{R_{\mathrm{vir}}}
\newcommand{\Cvir}{C_{\mathrm{vir}}}
\newcommand{\Reff}{R_{\mathrm{e}}}
\newcommand{\Mstar}{M_{\star}}
\newcommand{\etatherm}{\eta_{\mathrm{therm}}}
\newcommand{\PDF}{\mathit{P}}
\newcommand{\PzM}{\PDF(z_{f} | \log \MBH)}
\newcommand{\IPzM}{\PDF(z_{f}>z| \log \MBH)}
\newcommand{\qexp}{\left<q | \log \MBH \right>}

\shorttitle{$\msigma$ vs. Redshift}
\shortauthors{Robertson et al.}

\begin{document}

\title{The Evolution of the $\MBH$-$\sigma$ Relation}
\author{Brant Robertson\altaffilmark{1,5},
	Lars Hernquist\altaffilmark{1},
	Thomas J. Cox\altaffilmark{1},\\
        Tiziana Di Matteo\altaffilmark{2},
	Philip F. Hopkins\altaffilmark{1},
        Paul Martini\altaffilmark{3},
        Volker Springel\altaffilmark{4}}

\altaffiltext{1}{Harvard-Smithsonian Center for Astrophysics, 
        60 Garden St., Cambridge, MA 02138, USA}
\altaffiltext{2}{Carnegie-Mellon University, Department of Physics,
        5000 Forbes Ave., Pittsburgh, PA 15213, USA}
\altaffiltext{3}{The Ohio State University, Department of Astronomy,
        140 West 18th Ave., Columbus, OH 43210, USA.}
\altaffiltext{4}{Max-Planck-Institut f\"ur Astrophysik, Karl-Schwarzschild-Stra\ss e  1, 
	85740 Garching bei M\"unchen, Germany}
\altaffiltext{5}{brobertson@cfa.harvard.edu}

\begin{abstract}

We examine the evolution of the black hole mass - stellar velocity
dispersion ($\msigma$) relation over cosmic time using simulations of
galaxy mergers that include feedback from supermassive black hole
growth.
Our calculations have been shown to reproduce the
power-law scaling of the local, redshift zero $\msigma$ relation and
yield the observed normalization for an appropriate choice of the
feedback efficiency.
In the present paper, we consider mergers of galaxies varying the
properties of the progenitors to match those expected at various
cosmic times.  In particular, we examine a range in redshifts
$z=0$-$6$, where we modify the virial mass, gas fraction, interstellar
medium equation of state, surface mass density, and concentration of
the dark matter halos.  We find that the slope of the resulting
$\msigma$ relation is the same at all redshifts.  For the same
feedback efficiency that reproduces the observed amplitude of the
$\msigma$ relation at $z=0$, there is a weak redshift-dependence to
the normalization, corresponding to an evolution in the Faber-Jackson
relation, which results from an increasing velocity dispersion for a
given galactic stellar mass.
We develop a formalism to connect redshift evolution in the $\msigma$
relation to the scatter in the local relation at $z=0$.  For an
assumed model for the accumulation of black holes with different
masses over cosmic time, we show that the scatter in the local
relation places severe constraints on the redshift evolution of both
the normalization and slope of the $\msigma$ relation.  Furthermore,
we demonstrate that the cosmic downsizing of the black hole population
introduces a black hole mass-dependent dispersion in the $\msigma$
relation and that the skewness of the distribution about the locally
observed $\msigma$ relation is sensitive to redshift evolution in the
normalization and slope.  Improving the statistics of the local
relation to measure the moments of its distribution will therefore 
make it possible to
constrain the evolution of the $\msigma$ relation and black hole
population at redshifts that are currently not observationally
accessible.  In principle, these various diagnostics provide a method
for differentiating between theories for producing the $\msigma$
relation.
In agreement with existing constraints, our simulations imply that 
hierarchical structure formation should produce the relation with
small intrinsic scatter as the physical origin of the $\msigma$ 
enjoys a remarkable resiliency to the redshift-dependent properties 
of merger progenitors.

\end{abstract}

\keywords{galaxies: formation -- galaxies: evolution}

\section{Introduction}
\label{section:introduction}

Galactic spheroids exhibit a constitutive relation between their
stellar kinematics and supermassive black holes (SMBHs) through
observed correlations between black hole mass $\MBH$ and either the
bulge mass $\Mbulge$
\citep{miyoshi95a,kormendy95a,eckart97a,faber97a,magorrian98a,ghez98a},
or the velocity dispersion $\sigma$
\citep[$\msigma$,][]{ferrarese00a,gebhardt00a}.  The $\msigma$
relation locally is a tight, power-law correlation with logarithmic
slope originally estimated to be in the range $\beta \approx 3.75-4.8$
\citep{ferrarese00a,gebhardt00a}.  Later studies improved the accuracy
of the local correlation \citep[e.g.][hereafter T02]{tremaine02a} and explored
the relation between SMBH mass and other galaxy properties more fully
\citep{wandel02a,ferrarese02a, marconi03b,bernardi03a,baes03a}.  Efforts to extend
our knowledge of the connection between SMBHs and their host galaxies
to higher redshifts include attempts to measure SMBH properties at
$z=4-6$ \citep{vestergaard04a} and to quantify the relation between
SMBHs and galaxy kinematics in distant, active galaxies
\citep{shields03a, treu04a,walter04a}.

Observations of active galaxies at $z>0$ have yielded ambiguous
inferences about the nature of the $\msigma$ at other redshifts,
implying both redshift-dependent \citep{treu04a,walter04a} and
redshift-independent relations \citep{shields03a}.  Knowledge of the
$\msigma$ relation over cosmological time is crucial to understanding
galaxy formation if SMBH feedback has a significant impact on this
process \citep[e.g.][]{springel05a,robertson05a}.  The focus of our
present work is to characterize the development of the $\msigma$
relation over cosmic time with a large set of hydrodynamic
simulations of galaxy mergers that include star formation and feedback
from the growth of supermassive black holes, to use existing data to
constrain redshift evolution in the $\msigma$ relation, and to 
motivate specific observational tests for the $\msigma$ relation
at high redshifts using the next generation of large telescopes.

A variety of theoretical models have been proposed to explain the
$\msigma$ relation, based on physical mechanisms such as viscous disk
accretion \citep{burkert01a}, adiabatic black hole growth
\citep{macmillan02a}, gas or dark matter collapse
\citep{adams01a,balberg02a,adams03a}, stellar capture by accretion
disks \citep{miralda-escude05a}, dissipationless merging
\citep{ciotti01a,nipoti03a}, unregulated gas accretion
\citep{archibald02a,kazantzidis05a}, and the self-regulated growth of
black holes by momentum- or pressure-driven winds
\citep{silk98a,king03a, murray05a,di_matteo05a,sazonov05a}.  An
important outcome of the these studies has been the recognition that
the growth of SMBHs may have important consequences cosmologically,
especially for galaxy formation, motivating work linking the
hierarchical growth of structure with the evolution or demographics of
SMBHs through studies of quasar evolution \citep{
haiman98a,richstone98a,fabian99a,monaco00a,
kauffmann00a,haehnelt00a,granato01a,menou01a,menci03a,
di_matteo03a,di_matteo04a,wyithe03a,wyithe05a,hopkins05a,hopkins05b,hopkins05c,
hopkins05d}
and the formation of normal galaxies
\citep{volonteri03a,cox05a,robertson05a}.

Motivated by the long-standing view that mergers form spheroids
\citep[e.g.][]{toomre77a}, \cite{di_matteo05a} studied the
self-regulated growth of black holes in galaxy collisions and showed
that this process can reproduce the $\msigma$ relation at $z=0$ for
progenitors similar to local galaxies.
The work of \cite{di_matteo05a} served as an important early attempt
to calculate the effects of black hole growth and feedback on the
process of galaxy formation by addressing the generation of an 
$\msigma$ relation in self-consistent hydrodynamical simulations.
This calculation has served as a foundation for proceeding research
that has attempted to explore the emerging paradigm that supermassive
black hole formation is a natural and essential outcome of hierarchical
spheroid formation.
\cite{springel05a} calculated the role of AGN feedback in the reddening spheroids formed in mergers by truncating residual star formation.
\cite{robertson05a} explored the effects of AGN feedback in reducing the size of spheroids of disk galaxies forming in the early gas-rich environments of high redshifts.
\cite{cox05a} calculated the contribution of AGN outflows to the x-ray emission from halo gas in elliptical galaxies.
Importantly, \cite{hopkins05a,hopkins05b,hopkins05c,hopkins05d,hopkins05e,hopkins05f,hopkins05g} calculated a wide range of implications of the co-evolution of spheroids and black holes for the interpretation of both quasar and galaxy observations.
\cite{hopkins05a,hopkins05b,hopkins05c,hopkins05d} demonstrated that modeling the obscured growth of black holes during gas-rich galaxy mergers leads to a luminosity-dependent quasar lifetime that differs substantially from standard assumptions for the time-dependence of quasar activity.
The luminosity-dependent quasar lifetime leads naturally to a new interpretation of the quasar luminosity function, where low-luminosity quasars are primarily massive black holes in sub-Eddington accretion phases while quasars with luminosities above the break in the quasar luminosity function are accreting near the Eddington rate.
\cite{hopkins05e} show that the luminosity-dependent lifetime and the new interpretation of the quasar luminosity function can explain a wide variety of quasar observations by deconvolving the growing black hole population from the hard X-ray luminosity function, including longer-wavelength quasar luminosity functions, the spectrum of the X-ray background, and the obscured fraction of quasars as a function of quasar luminosity.
The model presented in \cite{hopkins05e} also leads to a new model for quasar clustering \citep{lidz05a} and predictions for the evolution of the faint-end slope of the quasar luminosity function \citep{hopkins05g}. 
Furthermore, \cite{hopkins05f} show that the co-evolution of spheroids and black holes in these hydrodynamical models can account for the evolution of the red galaxy sequence, building off the previous calculations of \cite{springel05a} and \cite{hopkins05e}.

Many of the previously mentioned calculations utilize knowledge of the connection between supermassive black holes and their host spheroids over cosmological time and throughout the process of hierarchical structure formation.  The calculations presented in the current paper provide a detailed hydrodynamical calculation of the nature of this
spheroid-black hole connection using cosmologically-motivated galaxy models at a variety of redshifts, and have already served as essential input into many the aforementioned calculations.
Independently, the detailed calculation of the $\msigma$ relation with redshift provides an important method for testing the premise of spheroidal
galaxy formation through the hierarchical merging of galaxies against
observational constraints from the $\msigma$ relation observed locally
and at high-redshifts.
These calculations will also serve as a basis for future comparisons between theoretical expectations for the co-evolution of black holes and spheroids with observations of ultra-luminous infrared galaxies \citep[e.g.][]{sanders88a,sanders96a} and the submillimeter emission from galaxies observed with SCUBA \citep[e.g.][]{smail97a}.

Specifically, in this paper we use the
\cite{di_matteo05a} model to explicitly calculate the behavior of the
$\msigma$ relation with cosmic time induced by the
redshift-scalings of galaxy properties.
Using a large set of
hydrodynamical simulations of mergers that include feedback from
supernovae and accreting SMBHs, the resulting $\msigma$ relation is
calculated for galaxy models that span a large range in virial mass,
gas fraction, interstellar medium equation of state, surface mass
density, dark matter concentration, and redshift.

We determine the allowed redshift evolution in either the normalization
or slope
of the $\msigma$ relation consistent with constraints
from the observed $\msigma$ relation at $z=0$ under the assumption 
that black holes of
mass $\MBH$ form at a characteristic epoch $z_{f}$.
The redshift
evolution of the normalization in our simulations yields a dispersion
in the $z=0$ relation consistent with the observed scatter.  We further relate
our simulations to ongoing efforts to measure the
$\msigma$ relation in high-redshift active galactic nuclei (AGN), and
discuss future observational tests for normal
galaxies at high redshifts with the next-generation of large aperture
telescopes.

We describe our simulation methodology in \S
\ref{section:methodology}, report our results in \S
\ref{section:results}, compare with observational constraints in \S
\ref{section:observations}, discuss our results in \S
\ref{section:discussion}, and summarize and conclude in \S
\ref{section:summary}.  For our cosmological parameters, we adopt a
flat universe with $\Omega_{\mathrm{m}}=0.3$, $\Omega_{\Lambda}=0.7$,
$\Omega_{\mathrm{b}}=0.04$, and $h=0.7$.

\begin{deluxetable*}{cccccc}
\tablewidth{0pt}
\tablecaption{\label{table:models}
}
\tablehead{
\colhead{Models} & \colhead{Progenitor $\Vvir$ [km s$^{-1}$]} & \colhead{Redshift $z$} & \colhead{Gas Fraction $f_{\rm gas}$} & \colhead{ISM Pressurization $q_{\rm EOS}$} & \colhead{Thermal Coupling $\eta_{\rm therm}$}}
\startdata
Local & $80$, $115$, $160$, $226$, $320$, $500$ & $0$ & $0.4$,  $0.8$ & $0.25$, $1.0$ & $0.05$ \\
Intermediate-$z$ & $80$, $115$, $160$, $226$, $320$, $500$ & $2$, $3$  & $0.4$, $0.8$ & $0.25$, $1.0$ & $0.05$ \\
High-$z$ & $115$, $160$, $226$, $320$, $500$ & $6$ & $0.4$, $0.8$ & $0.25$, $1.0$ & $0.05$ \\
Low $\eta_{\rm therm}$ & $115$, $160$, $226$, $320$, $500$ & $6$ & $0.4$, $0.8$ & $0.25$, $1.0$ & $0.025$ 
\enddata
\end{deluxetable*}

\section{Methodology}
\label{section:methodology}

We have performed a set of $112$ simulations of merging galaxies
using the entropy-conserving formulation \citep{springel02a}
of smoothed particle hydrodynamics \citep{lucy77a,gingold77a} as
implemented in the GADGET code \citep{springel01a, springel05c}.  Each
progenitor galaxy is constructed using the method described by
\cite{springel04a}, generalized to allow for the expected redshift
scaling of galaxy properties.  For a progenitor with virial velocity
$\Vvir$ at redshift $z$ we determine a virial mass and radius assuming
a virial overdensity $\Delta_{\vir} = 200$ times the critical density,
using the relations
\begin{eqnarray}
\label{equation:mvir}
\Mvir = \frac{\Vvir^{3}}{10GH(z)}\\
\label{equation:rvir}
\Rvir = \frac{\Vvir}{10H(z)}.
\end{eqnarray}
\noindent
For the Navarro-Frenk-White concentration \citep{navarro97a} of each
progenitor dark matter halo, we follow \cite{bullock01a} and adopt
\begin{equation}
\label{equation:cvir}
\Cvir \simeq 9 \left[\frac{\Mvir}{M_{\star}(z=0)}\right]^{-0.13} \left(1+z\right)^{-1}
\end{equation}
\noindent
where $M_{\star}(z=0) \sim 8 \times 10^{12} h^{-1} M_{\sun}$ is the
linear collapse mass at the present epoch and $H(z)$ is the Hubble
parameter.  The concentration and virial radius are then used to
create \cite{hernquist90a} profile dark matter halos using the
conversion described by \cite{springel04a}.  The halos are populated
with exponential disks of mass $M_{\rm disk} = 0.041 \Mvir$ and
fractional gas content $f_{\rm gas}$, whose scalelengths $R_{\rm d}$
are determined using the \cite{mo98a} formalism assuming the specific
angular momentum content $j_{\rm d}$ equals the disk mass fraction
$m_{\rm d}$ and a constant halo spin $\lambda = 0.033$.  The
distribution of halo spins in dissipationless simulations is measured
to be independent of mass and redshift \citep{vitvitska02a} and our
chosen value for $\lambda$ is near the mode of the distribution.  The
vertical scale heights of the stellar disks are set to $0.2 R_{\rm
d}$, similar to the Milky Way scaling \citep{siegel02a}.  The vertical
scale heights of the gaseous disks are determined via an integral
constraint on the surface mass density, a self-consistent
determination of the galaxy potential, and the effective equation of
state (EOS) of the multiphase interstellar medium
\citep[ISM,][]{mckee77a,springel03a}.
The model for the ISM utilized in these simulations accounts for the
multiphase nature of star-forming gas in galaxies by volume-averaging
the mass-weighted thermodynamic properties of cold clouds and a hot, diffuse 
component on scales not numerically resolved.
\cite{springel03a} present a detailed discussion of the implementation
and implications of this treatment of the ISM, but the primary dynamical effect
of this multiphase model is to pressurize the ISM as the hot, diffuse component
increases the effective temperature of the star-forming gas.
The effective equation of state of the gas \citep[for a numerical fit, see][]{robertson04a} therefore acts to stabilize dense gas in galactic disks against \cite{toomre64a} instabilities caused by self-gravitation \citep{springel03a,robertson04a,springel04a}.
The pressurization of the ISM
can be varied using an equation of state softening parameter $q_{\rm
EOS}$ \citep[for details, see][]{springel04a}, which linearly
interpolates between isothermal gas ($q_{\rm EOS} = 0$) and a strongly
pressurized multiphase ISM ($q_{\rm EOS} = 1$).

Each initial galaxy model consists of $40,000$ gas particles, $60,000$
dark matter particles, and $40,000$ stellar particles.  The
gravitational softening of gas and stellar particles is set to
$100h^{-1}(1+z)^{-1}$ pc.  The progenitors merge on prograde-parabolic
orbits, with an initial separation $R_{\rm init} = 5\Rvir/8$ and
pericentric distance $R_{\rm peri} = 2 R_{\rm d}$.  Each progenitor
galaxy contains a ``sink'' particle with a black hole seed of mass
$10^{5}h^{-1} M_{\sun}$ allowed to grow through a model based on
Bondi-Hoyle-Lyttleton
accretion \citep{hoyle39a,bondi44a,bondi52a}, implemented as a 
subresolution numerical model by \cite{springel04a}.  The Bondi 
accretion rate onto the black hole is determined by the surrounding
gas properties. To allow for black hole growth the BH particle is 
allowed to probabilistically accrete gas particles, but as explained 
in \cite{springel04a} this accreted gas is added to a reservoir 
from which the Bondi rate is smoothly calculated.  The BH particles
typically accrete $\sim100$ SPH particles during the evolution of
the simulations, depending on the details of the galaxy models and
mergers.
In this model, a
fraction $\epsilon_{\rm r} = 0.1$ of the mass accretion rate
$\dot{M}c^{2}$ is radiated and a fraction $\eta_{\rm therm}=0.05$ of
this $10\%$ is deposited as thermal feedback into the surrounding gas
in a kernel-weighted manner.
The thermal coupling is calibrated to reproduce the normalization of
the local $\msigma$ relation \citep{di_matteo05a}, and we note that
other prescriptions for the energy deposition from black hole feedback
can produce similar $\msigma$ scalings \citep[e.g.,][]{murray05a}.
We therefore expect our results to be somewhat insensitive to the
exact choice for how the feedback energy couples to the gas as long as 
the energy is deposited near the black hole.
As our intent is
to calculate the development of the $\msigma$ relation over time and
to avoid assumptions about the presence or form of the $\msigma$
relation at higher redshifts, the progenitor models do not contain
bulges but instead yield an $\msigma$ relation self-consistently
through the merger process.

\begin{figure*}
\figurenum{1}
\epsscale{0.8}
\plotone{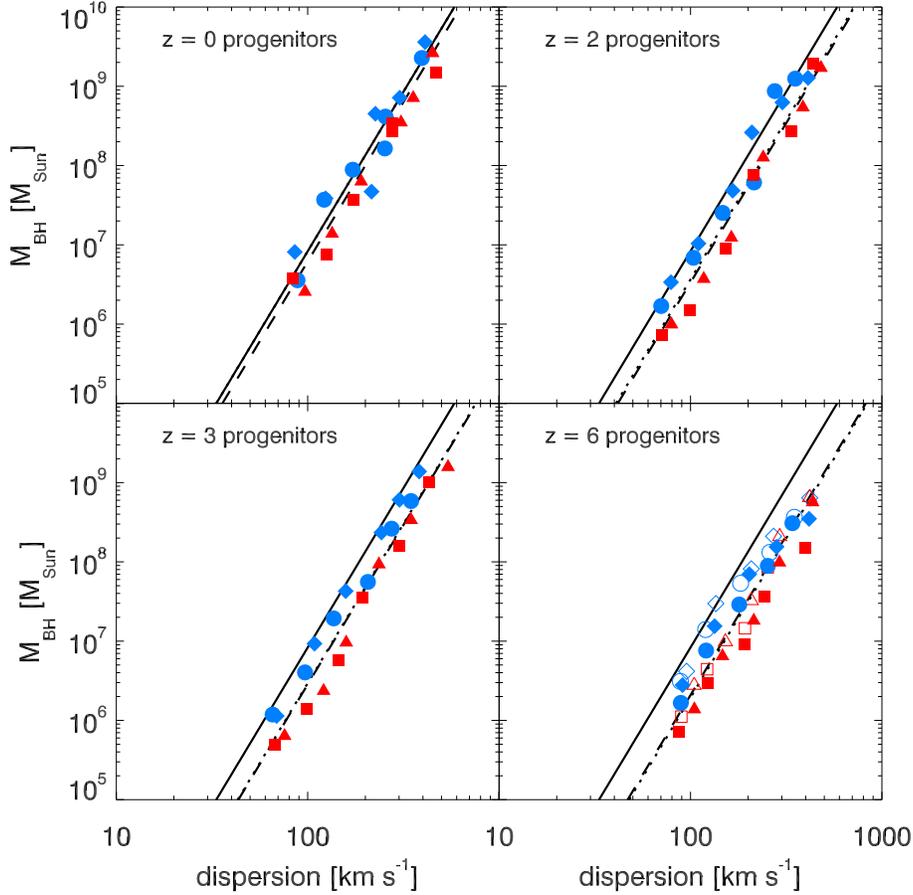}
\caption{\label{fig:m_sigma}
Relation between black hole mass $M_{\rm BH}$ and stellar
velocity dispersion $\sigma$ produced by the merging
of disk galaxy progenitors appropriate
for redshift $z=0$, $2$, $3$, and $6$.
These disk mergers produce nearly the same $\msigma$
scaling at all redshifts considered, while the 
simulated relations display a weak redshift evolution in their
normalization.  For normalization evolution of the form 
$\sigma \sim \sigma_{z=0}\left(1+z\right)^{\xi}$, the
normalization experiences a redshift scaling in the range
$\xi=0.138$ (relative to the simulated $z=0$ normalization,
dashed lines) to $\xi=0.186$ (relative to the observed $z=0$
normalization, dotted lines).
At each redshift we vary the gas fraction $f_{g}$ and
the pressurization of the interstellar medium $q_{\rm EOS}$
\citep[see][]{springel05a}, and we plot results for $M_{\rm BH}$ and
$\sigma$ in progenitor models with $g_{f}=0.4, q_{\rm EOS}=0.25$ (blue
circles), $g_{f}=0.8, q_{\rm EOS}=0.25$ (blue diamonds), $g_{f}=0.4,
q_{\rm EOS}=1.0$ (red squares), and $g_{f}=0.8, q_{\rm EOS}=1.0$ (red
triangles).
The \cite{tremaine02a} best fit to the local $\msigma$
relation (black line) is plotted for reference.
Also shown
is the $\msigma$ relation for $z=6$ progenitors calculated assuming a
decreased thermal coupling for black hole feedback (open symbols).
Lowering the thermal coupling for black hole feedback by half results
in only a small increase in the final black hole mass.
}
\end{figure*}

We simulate equal-mass mergers of progenitors appropriate for
redshifts $z=0$, $2$, $3$, and $6$ with virial velocities in the range
$\Vvir = 80-500$ km s$^{-1}$.  For each virial velocity, progenitors
with gas fractions $f_{\rm gas} = 0.4, 0.8$ and EOS softenings $q_{\rm
EOS} = 0.25, 1.0$ are simulated.  The $z=6$ runs are also repeated
using a thermal coupling constant of $\eta_{\rm therm}=0.025$ to gauge
the effect of altering the efficiency with which black hole feedback
heats the surrounding ISM.  Table \ref{table:models} summarizes our
full set of simulations.
The current set of simulations concerns the $\msigma$ relation generated
during the primary growth phase of the SMBH, which likely requires a major
merger to drive the massive inflow of gas to the central-most regions of
the colliding galaxies.  Future calculations of possible evolution in the
$\msigma$ relation should ideally involve cosmological simulations of galaxy
formation that naturally account for the mass spectrum of merger progenitors.
Although such simulations are still in development, we note that very minor 
mergers have been show to produce only weak star formation in the more massive
progenitor for even gas-rich systems as the feeble tidal torquing fails to
produce strong disk instabilities in the merging galaxies \citep{cox05b}.
The ability of the most minor mergers to affect a change in the $\msigma$
relation in the more massive progenitor galaxies should also be limited.

After the completion of a merger, the projected half-mass effective
radius $\Reff$ and the mass-weighted line-of-sight stellar velocity
dispersion $\sigma$ measured within an aperture of radius $\Reff$ are
calculated for each remnant galaxy for many viewing angles.  This
determination of $\sigma$ for simulated systems is commensurate with
the observational method of \cite{gebhardt00a} that measures a
luminosity-weighted velocity dispersion within an effective radius.
The black hole mass $\MBH$ in each galaxy is tracked during the
simulation and compared with the average dispersion $\sigma$ to
examine any resulting $\msigma$ correlation.

\begin{deluxetable}{cccccc}
\tablewidth{0pt}
\tablecaption{\label{table:fits} Best Fit $\msigma$ Relations
}
\tablehead{
\colhead{Progenitor Redshift} & \colhead{$\alpha$} & \colhead{$\beta$} & \colhead{$\Delta_{\log \MBH}$} & \colhead{$\alpha^{\beta=4.02}$} & \colhead{$\Delta_{\log \MBH}^{\beta=4.02}$}}
\startdata
$z=0$ & $8.01$ & $3.87$ & $0.26$ & $8.01$ & $0.26$\\
$z=2$ & $7.83$ & $4.10$ & $0.28$ & $7.83$ & $0.28$\\
$z=3$ & $7.72$ & $4.02$ & $0.26$ & $7.72$ & $0.26$\\
$z=6$ & $7.44$ & $3.62$ & $0.24$ & $7.45$ & $0.26$ 
\enddata
\end{deluxetable}

\section{Results}
\label{section:results}
Figure \ref{fig:m_sigma} shows the $\msigma$ relation produced by the
merging of progenitor galaxies appropriate for redshifts $z$ $=$ $0$,
$2$, $3$, and $6$ (upper left to bottom right).  At each redshift, we
perform a least-squares fit to the $\msigma$ relation of the form
\begin{equation}
\log \MBH = \alpha + \beta\log(\sigma/\sigma_{0})
\end{equation}
\noindent
where the relation is defined relative to $\sigma_{0} = 200$ km
s$^{-1}$. The observed relation has a normalization coefficient $\alpha = 8.13$ and slope $\beta
= 4.02$ (T02, solid line).  Table \ref{table:fits} lists the best fit
$\alpha$ and $\beta$ from the simulations at each redshift, along with
the dispersion $\Delta_{\log \MBH}$ about each best fit relation.
Table \ref{table:fits} also reports the best fit normalization
$\alpha^{\beta=4.02}$ and resultant dispersion $\Delta_{\log
\MBH}^{\beta=4.02}$ assuming the slope of the $\msigma$ relation is a
constant $\beta=4.02$ with redshift.  The intrinsic $\msigma$
dispersion at any given redshift induced by varying the gas fraction
or ISM equation of state is similar to the observed dispersion
$\Delta_{\log \MBH}=0.25-0.3$ (T02).  As Figure \ref{fig:m_sigma} and
Table \ref{table:fits} demonstrate, an $\msigma$ relation consistent
with a power-law scaling of index $\beta \simeq 4$ is predicted by our
modeling for normal galaxies out to high redshift.  We infer that the
physics that sets the scaling of the $\msigma$ relation is
insensitive to an extremely wide range of galaxy properties.  Virial
velocities of the progenitors range from $\Vvir=80$ km s$^{-1}$ to
$\Vvir=500$ km s$^{-1}$ at each redshift while the dark matter
concentrations range from $\Cvir\simeq15.6$ in the smallest progenitor
at $z=0$ to $\Cvir\simeq1.5$ in the largest progenitor at $z=6$.
For the $\msigma$ relation at each redshift, models with moderate ISM
pressurization typically experience slightly more black hole growth
than models with strong ISM pressurization.  
The increased thermal feedback in the model with a stiffer equation 
of state improves the dynamical stability of the gas, lessens gas 
angular momentum loss during the merger \citep{robertson04a,
robertson05a}, and results 
in less fueling of the black hole \citep{hopkins05e}.

\subsection{The Normalization of the $\msigma$ Relation with Redshift}

The normalization of the $\msigma$ relation evolves weakly with
redshift relative to the observed $z=0$ result.  To characterize this,
we generalize the parameterization of the $\msigma$ relation to
the form
\begin{equation}
\label{equation:xi}
\log \MBH = \alpha + \beta\log(\sigma/\sigma_{0}) - \xi\log(1+z)
\end{equation}
\noindent
with $\xi$ defined to be positive if the characteristic black hole
mass decreases with redshift or the characteristic velocity
dispersion increases with redshift.  
We determine the value
of $\xi$ for our simulations assuming either the observed $z=0$
normalization $\alpha_{\mathrm{obs}} = 8.13$ (Figure
\ref{fig:m_sigma}, dotted line) or the calculated $z=0$ normalization
$\alpha_{\mathrm{calc}}=8.01$ (Figure \ref{fig:m_sigma}, dashed line)
with the slope fixed at $\beta=4.02$.  We find evidence for a weak evolution, with
$\xi = 0.186$ for $\alpha_{\mathrm{obs}}$ and $\xi=0.138$ for
$\alpha_{\mathrm{calc}}$ using the best-fit parameters
reported in Table \ref{table:fits}.
We now consider whether the weak evolution we find in the
normalization of the $\msigma$ relation represents a physical effect
caused by the redshift-dependent properties of the progenitors or is
an artifact of our modeling.

\begin{figure}
\figurenum{2}
\epsscale{1.0}
\plotone{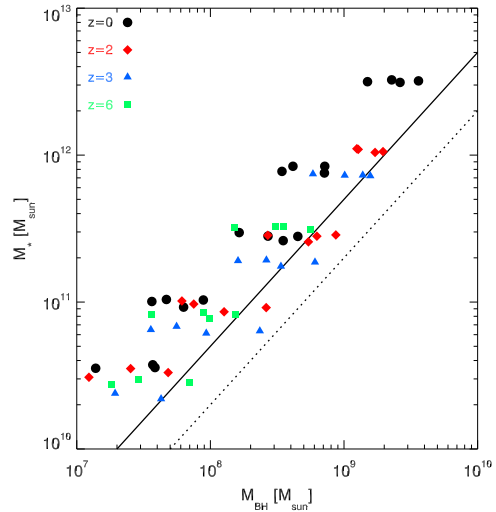}
\caption{\label{fig:bh_vs_mstar} Relation between total stellar mass $\Mstar$ and black hole mass $\MBH$ for merger progenitors appropriate for redshifts $z=0$ (black circles), $z=2$ (red diamonds), $z=3$ (blue triangles), and $z=6$ (green squares).
The ratio of black hole to stellar mass remains roughly constant for progenitors at all redshifts we consider, indicating that the normalization evolution of the $\msigma$ relation measured in the simulations is \emph{not} due to a decoupling of bulge and black hole growth.
At each circular velocity and redshift, four models for the gas fraction and interstellar medium equation of state are considered.  We plot bulge-dominated remnants only. The $\mbulge$ relations observed by \cite{marconi03b} (solid line) and \cite{magorrian98a} (dotted line) are shown for comparison.
}
\end{figure}

\subsubsection{Systematic Considerations}

The systematic effects of our modeling
include the redshift-scaling of halo virial properties and an assumed
fixed thermal coupling of black hole feedback to the gas in our
simulations.  The form of the progenitor galaxies is inferred from the
linear theory of dissipational galaxy formation and is therefore
idealized, but in the absence of cosmological simulations with
comparable resolution that include feedback from SMBH growth, our
choice appears well-motivated.  While the model considered here
assumes a fixed thermal coupling for black hole feedback, if the
thermal coupling changes with some bulk property of galaxies with
redshift (metallicity, for example) then this assumption may influence
our results.  The magnitude of thermal coupling $\etatherm=0.05$ of
black hole feedback to the surrounding gas is calibrated to reproduce
the locally observed relation.  To illustrate the impact of this
choice, Figure \ref{fig:m_sigma} shows the change in normalization of
the $\msigma$ relation for $z=6$ progenitors introduced by lowering
the thermal coupling parameter to $\etatherm=0.025$ (open symbols,
lower right panel).
As the coupling of the feedback to the gas of the galaxy is decreased, the
rate of energy input to the surrounding material by a black hole of a given mass and accretion rate is lessened.
The transition between Eddington-limited and self-regulated growth then occurs at a correspondingly higher black hole mass and, as a result, the final mass of the SMBH increases.
Decreasing the thermal coupling also mildly flattens the slope of the relation for the highest mass black holes.
Overall, the
change in the thermal coupling does not re-normalize the $\msigma$
relation for $z=6$ progenitors to coincide with the T02 result.  We
also note that varying the gas fraction in the progenitors from
$f_{\mathrm{gas}}=0.4$ to $f_{\mathrm{gas}}=0.8$ has little effect on
either the normalization or the scaling of the relation at any given
redshift.

A sensible measure to gauge a possible evolution of the black hole
growth with redshift is the comparison of black hole mass $\MBH$ with
total stellar mass $\Mstar$, which serves as a convenient proxy for
the observed $\mbulge$ relation.  Figure \ref{fig:bh_vs_mstar} shows
the $\MBH-\Mstar$ relation for mergers at all redshifts we consider,
with the \cite{marconi03b} (solid line) and \cite{magorrian98a}
(dotted line) $\mbulge$ relations plotted for comparison.  We show the
largest stellar mass systems, which are mostly spheroidal and lie near
the \cite{marconi03b} relation.
Any redshift-dependent impact on the black hole mass at a given
stellar mass is not discernibly
stronger than the black hole mass
increase for decreasing ISM pressurization and increasing
gas fraction.
We therefore conclude that the systematic
uncertainties of our modeling do not generate an unphysical evolution
in the black hole mass.
Instead, it appears more likely that the variation of the normalization of the $\msigma$ relation with redshift has a physical origin in the structural changes of the merging galaxies, a possibility we examine next.

\begin{figure}
\figurenum{3}
\epsscale{1.0}
\plotone{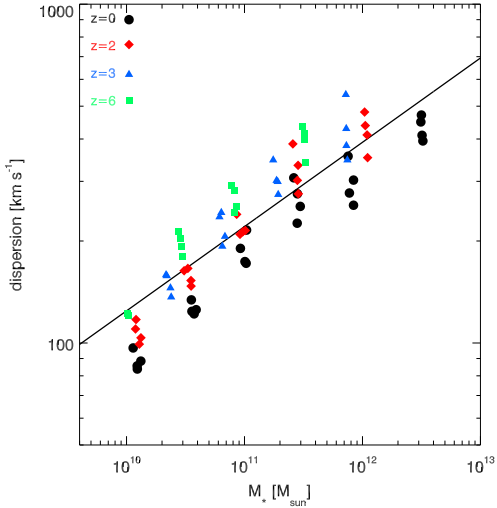}
\caption{\label{fig:sigma_vs_mstar} Relation between stellar velocity dispersion $\sigma$ and total stellar mass $\Mstar$ of remnants for merger progenitors appropriate for redshifts $z=0$ (black circles), $z=2$ (red diamonds), $z=3$ (blue triangles), and $z=6$ (green squares).
The simulated remnants display an increasing velocity dispersion for a given stellar mass with redshift,
which may be viewed as a weak evolution in the \cite{faber76a} relation. 
The evolution in velocity dispersion for a given stellar mass induces a corresponding evolution
in the calculated $\msigma$ relation (see Figure \ref{fig:m_sigma}).
At each circular velocity and redshift, four models for the gas fraction and interstellar medium equation of state are considered.
A total stellar-mass version of the Faber-Jackson relation constructed by combining the \cite{tremaine02a} $\msigma$ and \cite{marconi03b} $\mbulge$ relations is shown for comparison (solid line).
}
\end{figure}

\subsubsection{Physical Considerations}

The redshift-dependent structure of
progenitor systems expected from the cosmological scalings of
galaxy properties provides a plausible explanation for
the weak evolution in the normalization of the
$\msigma$ relation with redshift.
For example, the depth of the potential well in the center of each
remnant will influence the velocity dispersion, while for the regions
where the baryons dominate the stellar mass of the spheroid
determines the potential.  The characteristic densities of the
progenitors and remnants increase with redshift and any relation
involving the stellar mass may evolve cosmologically.  Similar
trends can be measured with our simulations.  Figure
\ref{fig:sigma_vs_mstar} shows the stellar velocity dispersion as a
function of the total stellar mass of each remnant for the simulations
at each redshift, along with a stellar mass formulation of
the Faber-Jackson luminosity-stellar velocity dispersion relation
found from the combination of the \cite{marconi03b} $\mbulge$ and T02
$\msigma$ relations (solid line).  For a given stellar mass, the
velocity dispersion increases for higher redshift progenitors, with
the greatest increase for the most spheroid-dominated systems.  We
therefore ascribe the weak evolution in the normalization of the
$\msigma$ relation with redshift to a systematically increasing
$\sigma$ for a given $\MBH$ or $\Mstar$, and argue that the steeper
potential wells of high redshift galaxies is a primary cause for this
apparent redshift evolution in the stellar mass formulation of the
Faber-Jackson relation.

The evolution of the stellar velocity dispersion for a given stellar
mass between redshifts $z=2$ and $z=6$ can be compared with the
redshift-dependent properties of our progenitor models.
Changes in the structure of galaxies can lead to an
evolution in the stellar velocity dispersion that scales with redshift
in a similar fashion to the evolving normalization in our calculated
$\msigma$ relations at redshifts $z=2-6$.
A possible origin for these effects can be illustrated
by the use of a simple galaxy model that demonstrates the impact of
redshift-dependent structural properties on the central stellar
kinematics.
To demonstrate this,
we consider the velocity dispersion of a stellar spheroid residing
within a dark matter halo with fixed virial mass $\Mvir$ as a function
of redshift.  For simplicity, we assume that the stellar mass within
the halo is a fixed fraction of the virial mass and that the central
potential is dominated by the baryons.  We model both the dark
matter halo and stellar spheroid as Hernquist profiles with
the stellar spheroid scalelength $a_{\mathrm{s}}$ a fraction $b$ of the
halo scalelength $a_{\mathrm{h}}$.  Equations
(\ref{equation:mvir}--\ref{equation:cvir}) describe the redshift
scalings of the dark matter halos for a given virial velocity
$\Vvir$.  Incorporating the results from \cite{springel04a}, we find
the spheroid scalelength will depend on virial mass and redshift as
\begin{eqnarray}
a_{\mathrm{s}}(\Mvir,z) &=& \frac{b}{\Cvir}\sqrt{2\left[\ln(1+\Cvir) - \Cvir/(1+\Cvir)\right]} \\ \nonumber
&\times& \left[G\Mvir\right]^{1/3}\left[10H(z)\right]^{-2/3}
\end{eqnarray}
\noindent
where $\Cvir$ is the mass- and redshift-dependent concentration given
by Equation (\ref{equation:cvir}).  The stellar velocity
dispersion of the Hernquist profile depends on the scalelength as
$\sigma \propto a_{\mathrm{s}}^{-1/2}$.  The power-law scaling of
$\sigma \propto (1+z)^{\xi}$ for $2<z<6$ then ranges from
$\xi\approx0.17$ for a $\Mvir \sim 10^{10} M_{\sun}$ halo to
$\xi\approx0.28$ for $\Mvir \sim 10^{14} M_{\sun}$, which encompasses
the approximate scaling $\xi$ found in the simulations.  Below
redshift $z=2$, the redshift evolution of the stellar velocity
dispersion begins to flatten and is roughly constant between redshifts
$z=0$ and $z=1$.  The relative constancy of the stellar velocity
dispersion in this redshift regime is caused by the growing
influence of the cosmological constant on the Hubble parameter as it
balances the increase in the virial concentration.  We therefore
expect the normalization of the $\msigma$ relation to remain roughly
fixed for purely dissipationless systems between $z=0$ and $z=1$, and
begin to evolve roughly as a weak power-law in $1+z$ toward higher
redshifts.  For the dissipational systems we simulate, an additional
increase in the stellar velocity dispersion between redshifts $z=0$
and $z=2$ beyond the estimate for purely stellar systems is needed to
fully explain the evolution measured in the simulations by redshift
$z=2$.  We leave such an exploration for future work, but note here
that either gas dissipation or feedback effects from the supermassive
black hole may influence the scaling of the stellar dispersion with
redshift or its mass dependence.

\begin{figure}
\figurenum{4}
\epsscale{1.0}
\plotone{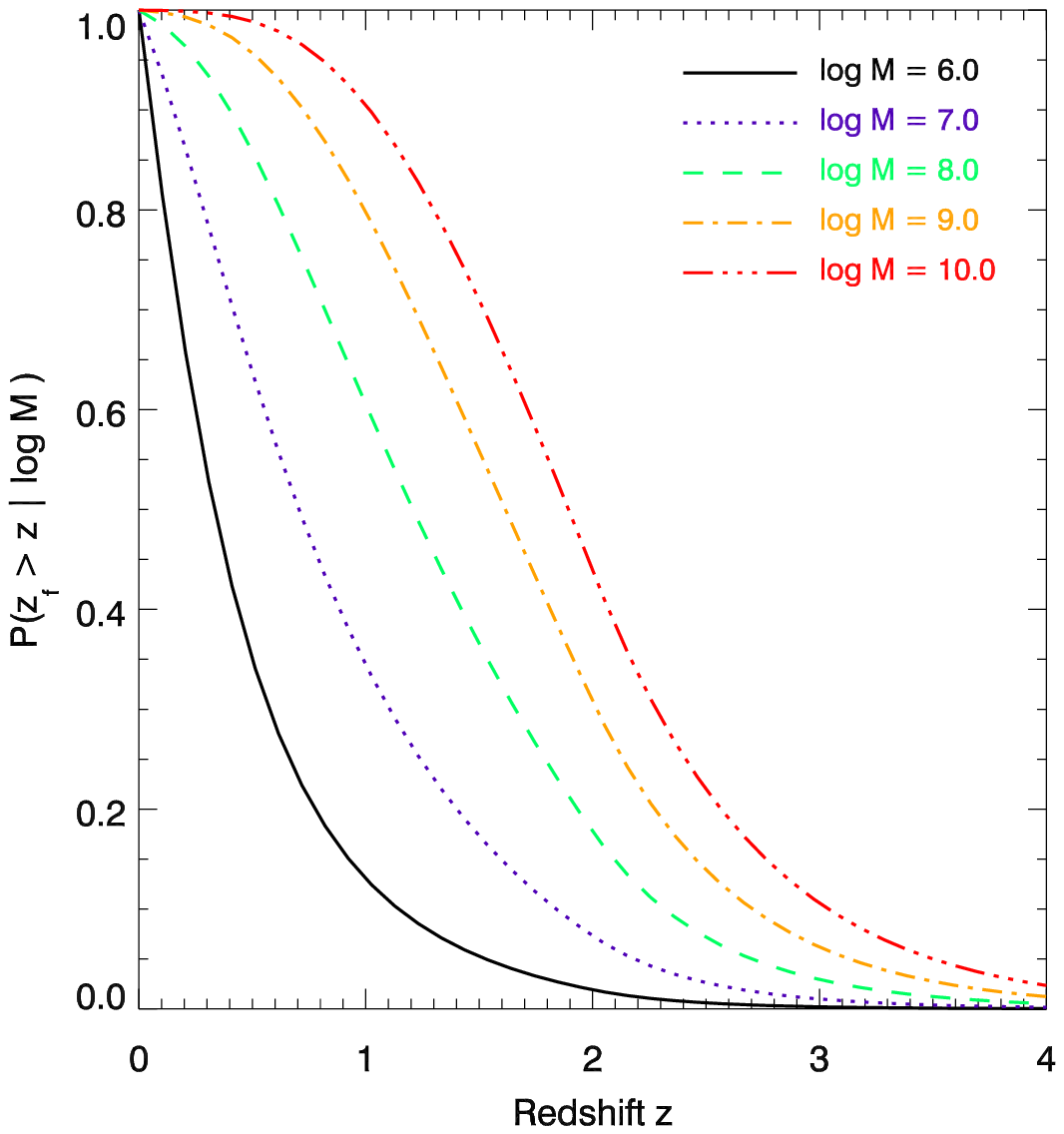}
\caption{\label{fig:PzM} Integrated probability distribution function $\IPzM$ for
a supermassive black hole of mass $\MBH$ to have formed at a redshift $z_{f}>z$,
as determined by the best fit model of \cite{hopkins05e}
(see \S \ref{sec:scatter} for details).
Shown is the $\IPzM$ for $\log \MBH = 6.0$ (solid black), $7.0$ (purple dotted), $8.0$ (blue dashed), $9.0$ (green dash-dotted), and $10.0$ (red dash-double dotted).
More massive black holes are typically formed at successively higher redshifts, reflecting the cosmic downsizing trend inferred from evolution in the quasar luminosity function.
This $\IPzM$ distribution is combined with functional forms for redshift-dependent $\msigma$ relations to estimate
constraints on possible evolution in the $\msigma$ relation by comparing with observations at $z=0$ (see Figures \ref{fig:mu_xi} and \ref{fig:mu_phi}).
}
\end{figure}

\section{Comparison with Observations}
\label{section:observations}

\subsection{Scatter Induced in the Local $\msigma$ \\Relation by Redshift Evolution}
\label{sec:scatter}

If the $\msigma$ relation evolves with redshift, any redshift
evolution in the cosmic black hole population will affect the
statistical distribution of galaxies about the observed $\msigma$
relation at $z=0$.  Following the formalism presented by
\citet[][YL04]{yu04a}, we can statistically characterize the
connection between active galactic nuclei and the local $\msigma$
relation by assuming BH growth occurs primarily during periods of
quasar activity.  If we further assume that the $\msigma$ relation
generated by mergers at each redshift has an intrinsically small
scatter, we can calculate the allowed redshift evolution in $\sigma$
given $\MBH$ consistent with the dispersion measured in the local
$\msigma$ relation.

From $\PzM$, the probability distribution function (PDF) for a black hole of mass $\MBH$ to have formed at a redshift $z_{f}$, we can calculate the expectation value of a function $q(z_{f},\log \MBH)$ given $\log \MBH$ as
\begin{equation}
\label{equation:mean}
\qexp = \int q(z_{f},\log \MBH) \PzM \mathrm{d} z_{f}
\end{equation}
\noindent
and the moments about $\qexp$ as
\begin{eqnarray}
\label{equation:moment}
\mu_{n}^{q} = \int \left[q(z_{f},\log \MBH) - \qexp\right]^{n}\\ \nonumber
	\times \PzM \mathrm{d} z_{f}.
\end{eqnarray}
We can identify the expectation value $\left<\sigma | \log \MBH\right>$ with the $\msigma$ relation observed at $z=0$, and associate the second moment with the measured $\msigma$ relation dispersion $\Delta_{\log \MBH}^{\mathrm{obs}} \equiv \sqrt{\mu_{2}^{\sigma}} = 0.25-0.3$ dex (T02).
If the velocity dispersion depends on some larger set of independent variables $\bar{\mathbf{x}}$ as $\sigma(\log \MBH, \bar{\mathbf{x}})$, and $\PzM$ is independent of $\bar{\mathbf{x}}$, we can extend this definition of the expectation value to calculate, e.g., $\left< \sigma | \log \MBH, \bar{\mathbf{x}}\right>$.
The analysis of YL04 treats a more general scenario where joint PDFs for properties associated with black hole mass are functions of more than one variable, and we restrict our attention to a single PDF directly relating $z_{f}$ and $\log \MBH$.
Equivalently, that $\PzM$ is independent of $\bar{\mathbf{x}}$ reflects our operating assumptions including: $1)$ all scatter in the $\msigma$ at $z=0$ owes to evolution in $\sigma$ with redshift (i.e. if $\bar{\mathbf{x}}$ is null, $\sigma$ does not depend on $z$ and the dispersion in the $\msigma$ relation would be $\Delta_{\log \MBH}=0$) and $2)$ galactic evolution subsequent to the primary growth phase of the black hole does not affect the direct connection between $z_{f}$ and $\log \MBH$.

The $\msigma$ relation generated by the merging of progenitor galaxies appropriate for a range of redshifts demonstrates an evolution in its normalization with respect to the observed relation at $z=0$.
By utilizing a known $\PzM$ and a functional form for $\sigma(z,\log \MBH, \bar{\mathbf{x}})$ that captures this evolution, the dispersion about the average relation $\left<\sigma | \log \MBH, \bar{\mathbf{x}} \right>$ can be calculated and compared with the observed $\Delta_{\log \MBH}^{\mathrm{obs}}$.
The small observed scatter in the $\msigma$ relation will then constrain $\bar{\mathbf{x}}$ for a known $\PzM$.
To this end, we consider $\msigma$ relations with an evolution in normalization of the form of Equation (\ref{equation:xi})
with the identification of $\bar{\mathbf{x}}=\xi$, or an evolution in the slope of the form
\begin{equation}
\label{equation:phi}
\log \MBH = \alpha + \beta(1+z)^{\phi}\log(\sigma/\sigma_{0})
\end{equation}
with $\bar{\mathbf{x}}=\phi$.
We proceed to calculate the resulting variance $\mu_{2}^{\sigma}$ about the mean $\msigma$ relation at $z=0$,
using these forms for $\sigma(\MBH,z)$ and the observed values of $\alpha=4.0$, $\beta=8.13$ (T02).
As the sign of either $\phi$ or $\xi$ will strongly affect the symmetry of the inferred distribution about the mean relation, we also characterize the  skewness $\gamma_{1} = \mu_{3}^{\sigma}/(\mu_{2}^{\sigma})^{3/2}$.

We now adopt a form for $\PzM$, using the results
of \citet[][H05]{hopkins05e}, and calculate the probability $\PzM$
for a black hole of mass $\MBH$ to have formed at redshift $z_{f}$.
The H05 modeling uses luminosity-dependent quasar lifetimes and
obscuring column density distributions measured from the simulations
presented in this paper to infer the black hole mass function $n(\MBH,z)$
from a host of constraints from observations of quasars.  The
evolution of $n(\MBH,z)$ with redshift is primarily determined by an
evolving characteristic luminosity $L_{\star}(z)$ that determines the
location of the maximum of $n(\MBH(L),z)$.  In principle, $\PzM$ is well
determined in the context of the H05 model for the cosmic black hole
population via the observed evolution of quasar luminosity functions
in multiple wavebands and constraints on the mass density of black
holes at $z=0$.  In practice, the somewhat weak observational
constraint on the evolution of the break in the luminosity function at
the highest observationally-accessible redshifts requires a choice for
the functional form of the evolving characteristic luminosity
$L_{\star}(z)$ that determines the peak of the black hole mass
function $n\left(\MBH(L),z\right)$.  As an \emph{Ansatz} we adopt the
form for $L_{\star}(z)$ chosen by H05 to best fit the available
constraints from observations, though we note that the calculations
presented in section \S 6 of H05 and below can be applied to more
general forms of $n(\MBH,z)$ and $L_{\star}(z)$ and the analysis
presented here could be repeated for other interpretations of the
quasar luminosity function than that adopted by H05.

The resulting $\PzM$, plotted in an integral form in Figure
\ref{fig:PzM} and reproduced from H05, reflects the best fit model for
the evolution of the cosmic black hole population given the available
data and the H05 interpretation of the quasar
luminosity function.  Cosmic downsizing, the concept that black holes of
larger mass formed characteristically at higher redshifts
\citep[e.g.][]{cowie03a,steffen03a}, is apparent in our $\PzM$ as the
formation-redshift PDF for low mass black holes increases strongly at
low redshifts.  The characteristic downsizing of black hole masses as
the universe ages leads to an immediate general conclusion about
possible evolution in the $\msigma$ relation: if black holes of
differing masses form at different characteristic redshifts, any
evolution in the $\msigma$ relation will introduce an $\MBH$-dependent
dispersion in the observed relation at $z=0$.  While subsequent
processes may reduce or amplify the mass dependence of $\Delta_{\log
\MBH}$, if observations reveal a trend in $\Delta_{\log \MBH}$ between
differing mass bins the result could be associated with an evolution
in the $\msigma$ relation.  Below, we characterize the mass dependence
of the statistical moments of the distribution of galaxies around the
mean $\msigma$ relation and relate them to possible forms for
$\msigma$ evolution.

\begin{figure*}
\figurenum{5}
\epsscale{1.1}
\plottwo{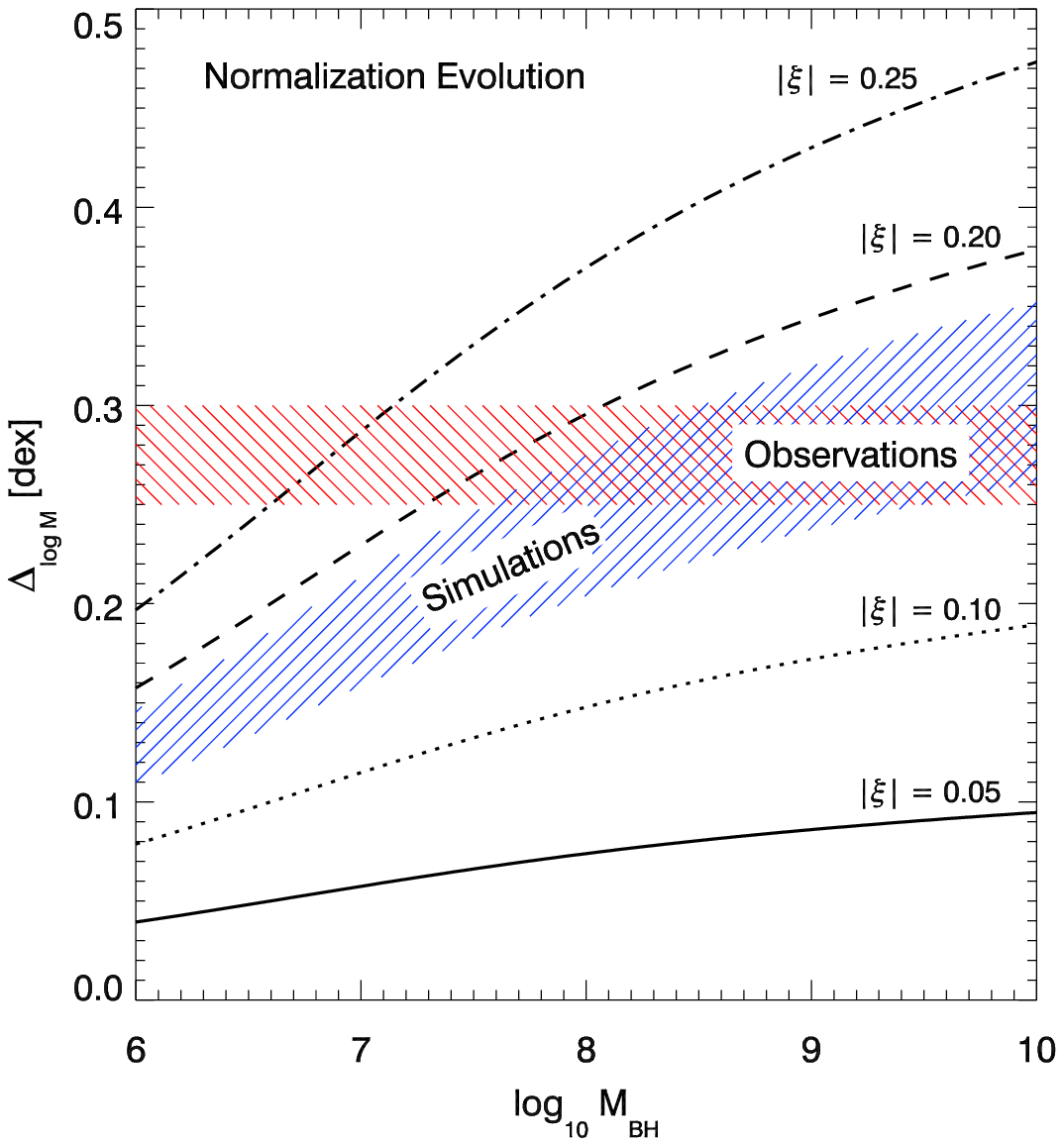}{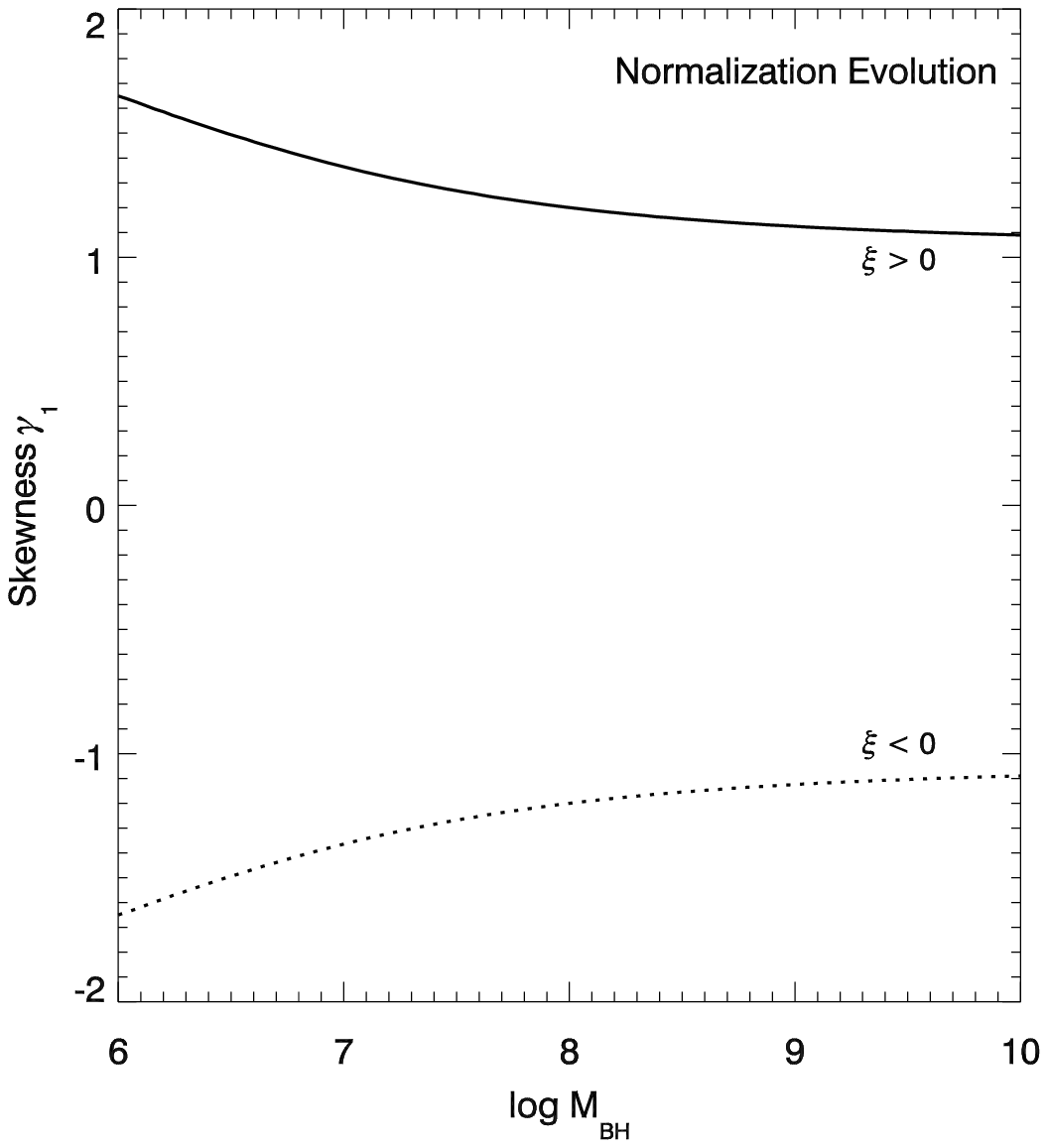}
\caption{\label{fig:mu_xi}
Dispersion $\Delta_{\log \MBH}$ and skewness $\gamma_{1}$ about the local $\msigma$ relation caused by evolution in the normalization of the $\msigma$ relation with redshift of the form $\log(\MBH) = \alpha + \beta \log(\sigma/\sigma_{0}) - \xi\log(1+z)$.
Redshift-evolution in the normalization induces a dispersion in the $\msigma$ relation at $z=0$ since black holes of 
different masses form at characteristically different cosmological times (see Figure \ref{fig:PzM}).
The dispersion induced by normalization evolution with redshift-scalings of $|\xi| = 0.05$ (solid), $0.10$ (dotted), $0.20$ (dashed), and $0.25$ (dash-dotted) is plotted.
The observed dispersion $\Delta_{\log \MBH}^{\mathrm{obs}} = 0.25-0.3$ (shaded region) provides a tight constraint 
on the evolution in the normalization, which is limited to approximately $|\xi| < 0.2$.
If the normalization evolves, the sign of the skewness in the distribution about the local $\msigma$ relation indicates the sign of $\xi$.}
\end{figure*}

\subsubsection{Normalization Evolution}
\label{subsubsec:normalization_evolution}

For a normalization evolution of the form of Equation
(\ref{equation:xi}), the dispersion $\Delta_{\log \MBH}$ and skewness
$\gamma_{1}$ for the $\msigma$ relation at $z=0$ can be readily
calculated using Equations (\ref{equation:mean}--\ref{equation:moment})
given a $\PzM$ describing the evolution of the cosmic black hole
population.  Figure \ref{fig:mu_xi} shows the dispersion in black hole
mass $\Delta_{\log \MBH}$ as a function of SMBH mass introduced in the
$\msigma$ relation at $z=0$ by normalization evolution over a range in
the power law index $\xi$ (left panel).  As $\Delta_{\log \MBH}$ is an
even function of $\xi$, we plot $\Delta_{\log \MBH}$ for different
values of $|\xi|$.  The observed range of $\Delta_{\log
\MBH}^{\mathrm{obs}} = 0.25-0.3$ dex (red shaded region) can be
compared to the normalization evolution measured in the simulations of
$\xi\approx0.138-0.186$ (blue shaded region), and the simulations
agree well with the observational constraints.  Given the
observational data and our chosen $\PzM$, evolution in the
normalization with approximately $|\xi|< 0.2$ is allowed.  The
mass-dependence of $\Delta_{\log \MBH}$ directly reflects changes
in the black hole population during the evolution of the $\msigma$
normalization with redshift, with more massive black holes being
formed during characteristically earlier epochs when the $\msigma$
normalization was more discrepant from the $z=0$ relation than during 
the formation epoch of smaller mass black holes.
As the most massive
black holes form over a wider range in redshift than smaller black holes,
their broader PDFs also contribute to the scatter about the mean relation.
If further sources of
scatter in the $\msigma$ relation are unimportant and subsequent
galactic evolution does not strongly effect $\PzM$, then the slope of
$\Delta_{\log \MBH}$ provides direct information about the typical
redshift of formation for black holes of a given mass.  We note that
since the black hole and spheroids form contemporaneously in our
model, the evolution owing to $\xi$ in the calculations also suggests
a correlation between the deviation of a galaxy from the $\msigma$
relation and the age of the spheroidal stellar population.  For
normalization evolution such a correlation results from the monotonic
mapping of the measured velocity dispersion deviation $\delta_{\log
\sigma} \equiv [\log \sigma(\MBH,z_{f}) - \log \sigma(\MBH,z=0)] = \xi
\log(1+z_{f})$ to a formation redshift $z_{f}$, which means the
probability distribution function $\PDF(\delta_{\log \sigma} | \log
\MBH)$ maps monotonically onto $\PzM$.

Further information on $\xi$ can be gleaned from the skewness
$\gamma_{1}$ of the distribution (Figure \ref{fig:mu_xi}, right
panel).  For normalization evolution of the form of Equation
(\ref{equation:xi}), the resulting skewness has the same sign as the
power law index and immediately indicates whether the $\msigma$
normalization increases or decreases with redshift.  All positive
values of $\xi$ have the same skewness, with the same functional shape
as the skewness defined by negative values of $\xi$ reflected about
the origin.

\begin{figure*}
\figurenum{6}
\epsscale{1.1}
\plottwo{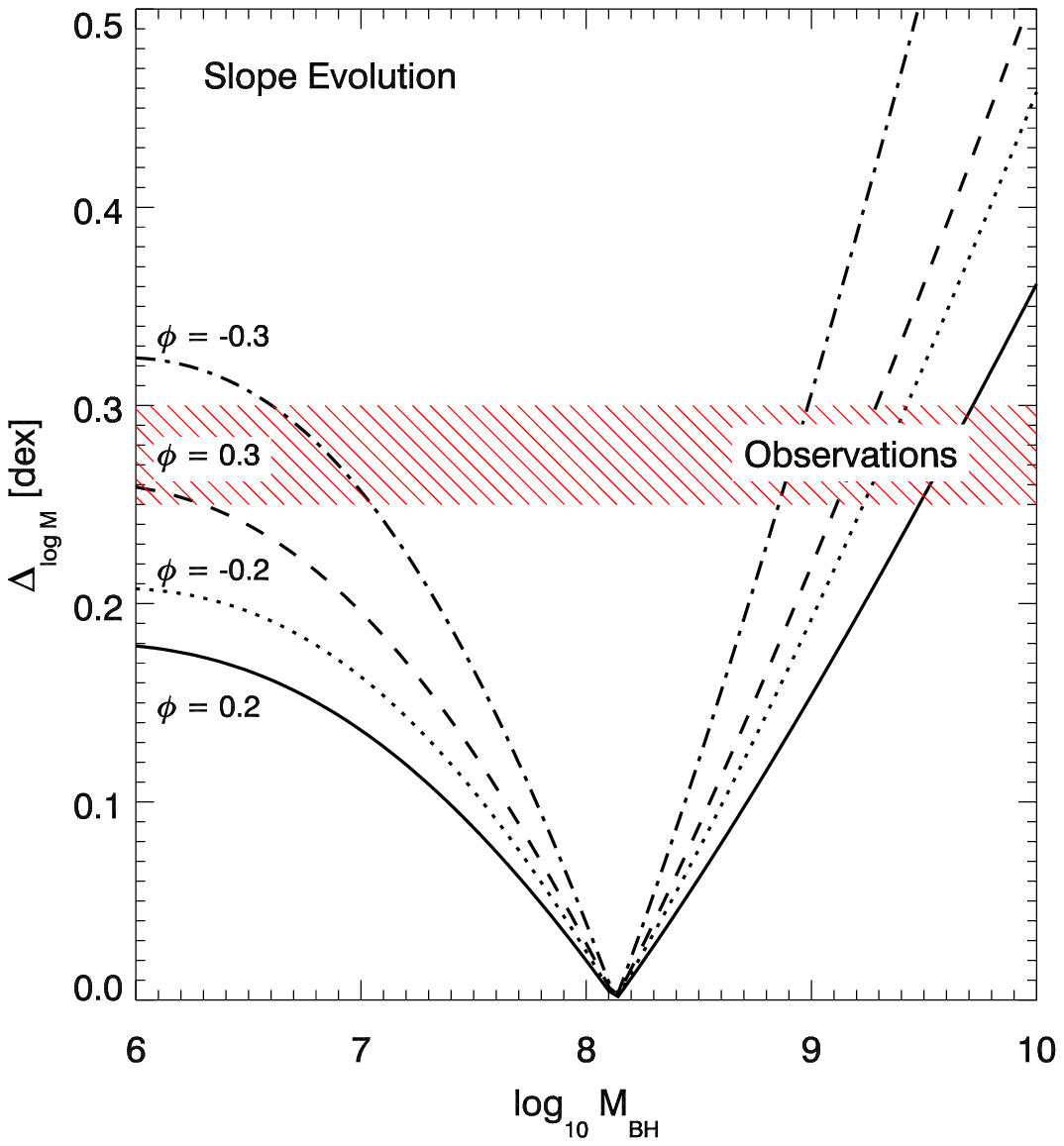}{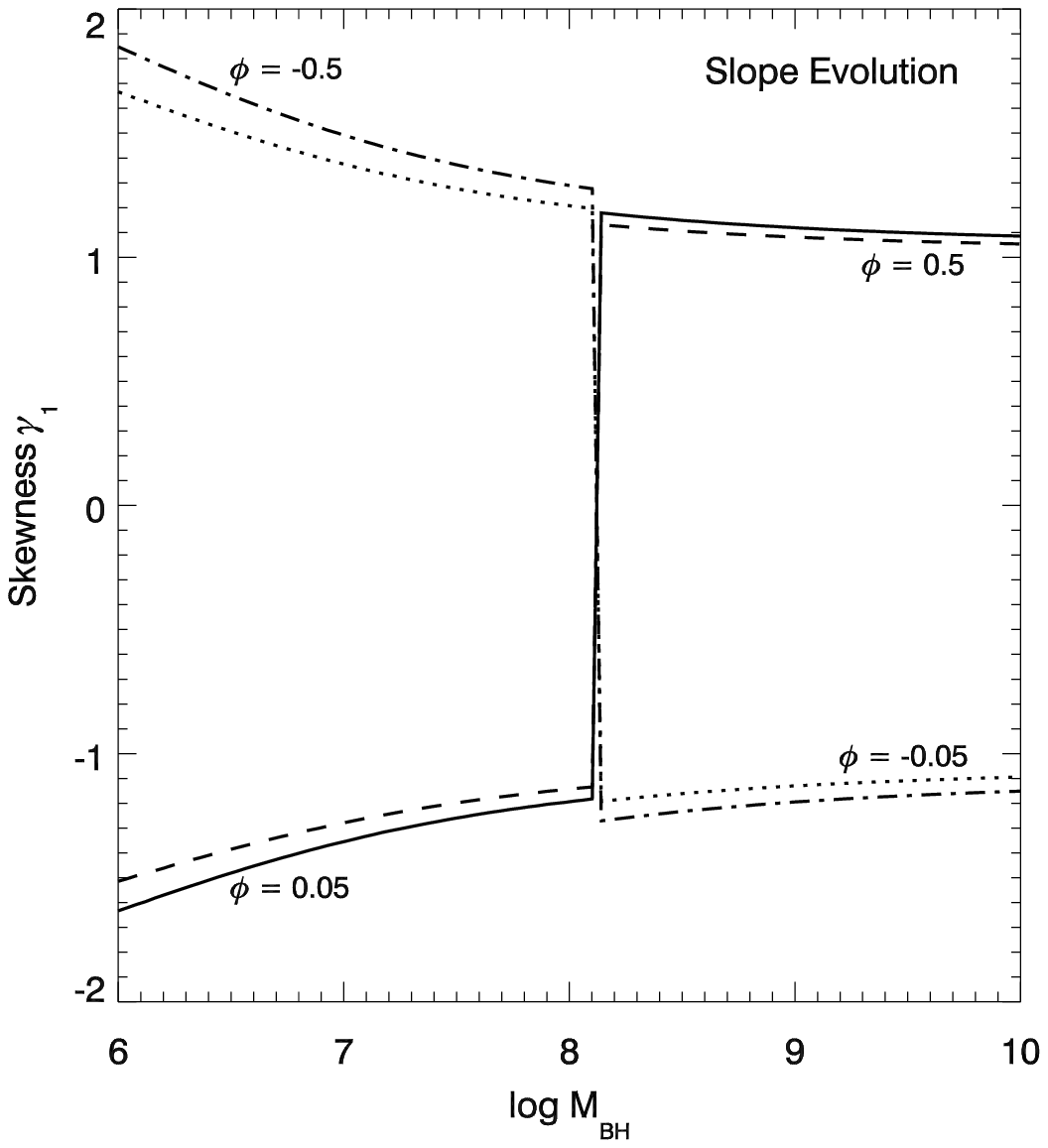}
\caption{\label{fig:mu_phi}
Dispersion $\Delta_{\log \MBH}$ (left panel) and skewness $\gamma_{1}$ (right panel) about the local $\msigma$ relation induced by an evolution in the slope of the $\msigma$ relation with redshift of the form $\log(\MBH) =  \alpha + \beta(1+z)^{\phi} \log(\sigma/\sigma_{0})$.
Redshift-evolution in the slope induces a dispersion in the $\msigma$ relation at $z=0$ since black holes of 
different masses form at characteristically different cosmological times (see Figure \ref{fig:PzM}).
The dispersion is plotted for $\phi = 0.2$ (solid), $0.3$ (dashed), $-0.2$ (dotted), and $-0.3$ (dash-dotted).
To illustrate the mass dependence of $\Delta_{\log \MBH}$ and $\gamma_{1}$ above and below the pivot, we choose the fixed intercept to be $\log M_{\mathrm{BH,fixed}} = \alpha = 8.13$.
The observed dispersion $\Delta_{\log \MBH}^{\mathrm{obs}} = 0.25-0.3$ (shaded region) constrains evolution in the slope to approximately $|\phi| < 0.3$.
If the slope of the $\msigma$ relation evolves and the intercept falls in the observable range, the sign of $\phi$ could be constrained by examining the sign of the observed skewness as a function of black hole mass because $\gamma_{1}$ changes sign at the pivot mass.
}
\end{figure*}

\subsubsection{Slope Evolution}
\label{subsubsec:slope_evolution}

While our simulated $\msigma$ relations do not show any monotonic
trend in their slope with redshift, many theories for the
$\msigma$ relation have some functional dependence in the slope with
redshift or black hole mass \citep[e.g.][]{sazonov05a,wyithe05a}.  The
dispersion $\Delta_{\log \MBH}$ and skewness $\gamma_{1}$ for the
$\msigma$ relation owing to slope evolution of the form of Equation
(\ref{equation:phi}) is also accessible using the method described for
characterizing normalization evolution.  Figure \ref{fig:mu_phi} shows
the dispersion $\Delta_{\log \MBH}$ (left panel) about the local
$\msigma$ relation induced by slope evolution, plotted for $\phi =
0.2$ (solid), $0.3$ (dashed), $-0.2$ (dotted), and $-0.3$
(dash-dotted).  To illustrate the mass dependence of $\Delta_{\log
\MBH}$ and $\gamma_{1}$ above and below the pivot we choose the fixed
intercept to be $\log M_{\mathrm{BH,fixed}} = \alpha = 8.13$, but the
same analysis could be performed for any intercept.  For consistency
with the observed dispersion $\Delta_{\log \MBH}^{\mathrm{obs}} =
0.25-0.3$ (shaded region), the evolution in the slope is constrained
to approximately $|\phi| < 0.3$ for our chosen intercept.  If the
fixed pivot intercept for a slope evolution of the $\msigma$ relation
is small ($\log \MBH \approx 6$), then the constraint on $|\phi|$
would be much tighter as $\Delta_{\log \MBH}$ is a strong function of
$\MBH$ above the pivot mass.  The skewness induced by a slope
evolution (Figure \ref{fig:mu_phi}, right panel) contains more
information than for normalization evolution.  The skewness provides a
constraint on the pivot mass if the pivot falls in the observationally
accessible range, as the sign of the skewness changes at this
intercept in the redshift-dependent relation.

In principle, any model that produces a redshift evolution in the
$\msigma$ relation can be compared with the observations in a similar
manner, and given a $\PzM$ for the cosmic black hole population,
distinct models for generating the $\msigma$ relation could be
observationally differentiated.  We note here that comparisons with
observations are most meaningful if the black holes in the
measured $\msigma$ relation represent a fair sample of the total black
hole population for a given mass bin.
For instance, an increase in the number of bulges in spiral systems 
with well-determined $\MBH$ and $\sigma$ would be helpful.  Unfortunately,
dust effects make stellar dynamical measurements late-type galaxies
difficult and a more effective approach for determining black hole
masses in these systems will need further development \citep[e.g.][]{ho02a}.
We encourage further
observational efforts to eliminate selection biases in the sample used
to determine the local $\msigma$ relation.

\subsection{Observations of $\msigma$ in AGN at $z>0$}
\label{subsec:high_z_agn}

Our simulations can be directly related to the $\msigma$ relation in normal galaxies at redshifts $z>0$.  However, previous observational efforts to measure the $\msigma$ relation in galaxies beyond the locally accessible sample have concentrated mostly on using the $\msigma$ relation indicators in active galactic nuclei \citep{shields03a,treu04a}, combining $\mathrm{H}\beta$ linewidth estimates for black hole mass \citep[e.g.][]{wandel99a} and [O{\sc iii}] linewidths as a proxy for stellar velocity dispersion \citep[e.g.][]{nelson00a} or dispersion measurements from stellar spectral features \citep{treu01a}.
Other estimates for the $\msigma$ relation in high-$z$ quasars have been made \citep[e.g.][]{walter04a} by comparing a black hole mass inferred from Eddington-limited accretion onto a bright quasar with the dynamical mass estimated from Very Large Array observations of molecular gas kinematics.

\cite{shields03a} found that AGN between $z=0-3.5$ obey an H$\beta$-[O{\sc iii}]
relation consistent with the local $\msigma$ relation determined by
T02, and the observations at any single redshift were also consistent
with the $z=0$ result, though the statistics are somewhat limited.
The observations of \cite{treu04a} imply a conflicting result with
seven AGN at $z\sim0.37$ inferred to primarily lie above the $z=0$
$\msigma$ relation, though with significant scatter and large
estimated error.  The \cite{walter04a} CO(3-2) observations also place
the $z=6.42$ quasar above the $z=0$ $\msigma$ relation and claim to
be inconsistent with the presence of a stellar bulge of the size
needed to match the local $\mbulge$ relations.

The increased scatter observed in the AGN samples and the deviation of
high-$z$ quasars relative to our calculations may originate in the
dynamical state of the observed systems \citep[for a related discussion, see][]{vestergaard04a}. 
If these systems correspond to merging
galaxies, they may not be dynamically relaxed and these samples may
therefore display a larger range of velocity dispersions than do the
dynamically relaxed systems that primarily comprise the locally
observed $\msigma$ relation or the redshift-dependent $\msigma$
relation in relaxed galaxies produced by our simulations.  
For instance, observations of local Seyfert galaxies show that
tidally-distorted hosts have systematically higher emission line
widths at a given stellar velocity dispersion \citep{nelson96a}.
An exploration of the $\msigma$ relation during the peak AGN phases of
our simulations is planned in future work, but given the current
observational and theoretical uncertainties we do not believe our
simulations are in conflict with the observations.  However, our
simulations are more consistent with the \cite{shields03a}
claim of no evolution than with either the \cite{treu04a} or \cite{walter04a}
claims of evidence for evolution in the $\msigma$ relation.  
In our model where the $\msigma$ scaling is set during
spheroid formation, the evolution of systems tends to occur either
quickly in the $\MBH$ direction during the exponential growth phase of
the black hole or more gradually along the $\msigma$ relation during
self-regulated black hole growth.  The presence of massive black holes
at high redshift without massive bulges would be rare in the
context of our modeling, and such outliers would likely have to
undergo purely stellar major mergers to reach the local $\msigma$
relation.  We also note that if the \cite{treu04a} 
offset of $\delta \log \sigma = -0.16$ measured for AGN
at $z=0.37$ holds for the $\msigma$
relation generated by the merging of normal galaxies at that redshift, then 
a power-law evolution in the normalization of $\sigma$ would imply 
$\xi=-1.17$.  If some subsequent process does not reduce the deviation
from the $z=0$ $\msigma$ relation for a given galaxy, then with our
$\PzM$ and the evolution $\xi=-1.17$ inferred from the \cite{treu04a}
measurements, the scatter in the $\msigma$ relation at $z=0$ would be
$\Delta_{\log \MBH} = 0.92-2.215$ over the range $\log \MBH =6-10$.
This scatter is much larger than both the observed dispersion
and the dispersion predicted by our simulations and formation
redshift distribution $\PzM$.

\subsection{Observations of Inactive Galaxies at $z>0$}
\label{subsec:high_z_galaxies}

For the next-generation of large telescopes, a measurement of the
central stellar velocity dispersion in normal galaxies at high
redshifts may be feasible, and our simulations can connect directly
with such observations.  A proper census of the cosmic black hole
population as a function of redshift will necessarily involve an
estimate of the mass of SMBHs at the centers of inactive galaxies.
Without black hole activity to provide an estimate of the black hole
mass, and given the likely prohibitively extreme resolution ($<10$ pc)
needed for stellar dynamical estimates of SMBH masses, inferring
black hole masses through the $\msigma$ relation via stellar velocity
dispersion measures for the cosmological spheroid population may
provide the best way of constraining the evolution of the black hole
population with redshift.
These measurements will necessarily involve an assumed model to
relate $\MBH$ and $\sigma$ as a function of redshift, such as the relation
calculated by our simulations.
Our work then provides additional
theoretical motivation for attempting these observations with future
generations of large infrared-capable telescopes.

Among the large telescopes currently in various stages of development are
the Giant Magellan Telescope \citep[GMT,][20-25m]{johns04a}, California
Extremely Large Telescope \citep[CELT,][30m]{nelson_j00a}, Euro50
\citep[][50m]{andersen04a}, OverWhelmingly Large Telescope
\citep[OWL,][100m]{dierickx04a}, Japanese Extremely Large Telescope
\citep[][30m]{iye04a}, and the Chinese Future Giant Telescope
\citep[][30m]{ding_qiang04a}.  The planned capabilities of these
telescopes are impressive.  For instance, the GMT design calls for a
resolution of $\theta_{\mathrm{GLAO}} \approx 0.15''$ with
ground layer adaptive optics (GLAO) and $\theta_{\mathrm{FAO}} \approx
0.007''$ with fully adaptive optics (FAO), corresponding to physical
scales at redshift $z=3$ of $R_{\mathrm{GLAO}} \approx 1.15$ kpc and
$R_{\mathrm{FAO}}\approx 54$ pc, respectively.  The remnant systems in
our simulations appropriate for $z=3$ progenitors have effective radii
ranging from $R_{\mathrm{eff}}\approx 0.3-10$ physical kpc, making
central stellar velocity dispersion measurements feasible in
principle, and likely possible even with GLAO observations for the
most massive systems.  Infrared spectroscopy of rest-frame UV stellar
lines could be used to find central stellar velocity dispersions of
spheroids and, combined with the simulation results to estimate $\MBH$
via an inversion of the calculated $\msigma$ relation for a given
redshift, the cosmic black hole population can be constrained
\citep[see, e.g.][]{yu02a}.
Future lensing observations with the
Square Kilometer Array to measure black hole
masses at intermediate redshifts \citep{rusin05a} may also be useful
in this regard.
In addition, if SMBH feedback is
important to determining the properties of normal galaxies
\citep[e.g.][]{springel04a, robertson05a}, such observations could
prove to be crucial input for theories of galaxy formation during the
evolutionary epoch when the most massive black holes are formed.

\section{Discussion}
\label{section:discussion}

Our simulations demonstrate that the merging of disk galaxies whose
properties are appropriate for redshifts $z=0-6$ should produce an
$\msigma$ relation with the observed power-law scaling $\MBH \propto
\sigma^{4}$.  The $\msigma$ relation retains this scaling at redshifts
$z=0-6$, even while the merger progenitors range widely in their
structural properties with redshift.  A possible weak evolution in the
normalization of the $\msigma$ relation is found, which may result
from an increasing velocity dispersion owing to the redshift evolution
of the potential wells of galaxies.  Simple arguments for how such
evolution would affect the velocity dispersion provide a similar
redshift scaling at redshifts $z=2-6$ to those calculated from the
simulations.  As these redshift scalings systematically place
high-redshift systems below the $z=0$ $\msigma$ relation, subsequent
accretion processes may further reduce the scatter in the observed
$\msigma$ relation over time.

Our results also provide guidance
to semi-analytical models about the behavior of the $\msigma$ relation.
We suggest that semi-analytical models in the vein of \cite{volonteri03a} or \cite{menci03a} could be altered to explicitly maintain the local $\mbulge$ relation at redshifts $z<6$, produce the same $\msigma$ scaling at redshifts $z<6$, and reproduce the local $\msigma$ normalization.
Simultaneously such models should tie morphological information, such as spheroid growth, to the growth of supermassive black holes \citep[for a related discussion, see][]{bromley04a}.
While some semi-analytical models include prescriptions for black hole accretion or feedback \citep[e.g.][]{menci03a,wyithe03a}, it is difficult to graft all the effects of feedback onto these models without a detailed treatment of the complex interplay between gas inflows, starbursts, and black hole activity.
These models could also be altered to more fully capture the black hole mass-dependent effects of feedback on galaxies and their environments \citep[see also, e.g.,][]{hopkins05e}.

Overall, our simulations provide more general conclusions about the
cosmological evolution of structure.  The spheroidal components of
galaxies are likely formed in the same process that sets the $\msigma$
relation and therefore the demographics of black holes traces the
demographics of spheroids over cosmic time, as is inferred locally
\citep{yu02a}.  With the possible shared origin of spheroids and SMBHs
in mind, we note that observations suggest the stellar populations of 
cluster ellipticals form at
redshifts $z>2$ \cite[e.g.][]{van_dokkum03a,gebhardt03a}.
This era coincides with the epoch $z\sim3$ 
during which the most massive SMBHs are
formed, as inferred from the evolution in the quasar number density
\citep{schmidt95a,boyle00a}.
Future comparisons with high-redshift AGN and massive ellipticals
could help provide
our modeling of individual mergers with a broader cosmological
context.  By connecting with the evolution of the observe quasar
luminosity function through the $\PzM$ supplied by calculations in
the vein of \cite{hopkins05e}, our model may account for cosmological 
considerations such as evolution in either the mass-dependent galaxy 
merger rate, the cold gas fraction, or the global star formation rate.

While uncertainties remain in our method, 
an analysis similar to that presented here could be
performed for simulations with different treatments of the relevant
physics of black hole accretion and feedback, star formation, or 
ISM processes.
As the observational data improve, the comparisons presented
in this paper could be used to constrain the physical
prescriptions used in simulations of galaxy formation.
For instance, our knowledge of possible redshift-dependence in the
thermal coupling $\etatherm$ of black hole feedback to the gas is limited.
The normalization of the $\msigma$ relation depends on $\etatherm$ for
a given accretion model, and a strong increase in $\etatherm$ would
introduce a correspondingly larger scatter in the $\msigma$ relation at
$z=0$.  A decrease in the thermal coupling with redshift would decrease
the scatter at $z=0$, but in principle this effect could be counteracted
if the dominant mode of black hole accretion is more efficient at higher
redshifts.  Better observational samples will help inform our modeling
of these physics by improving constraints from the statistical moments
of the $z=0$ $\msigma$ relation.

While our
method for comparing evolution in the $\msigma$ relation with
observational constraints involves adopting a probability
distribution function $\PzM$ relating black hole mass with a
characteristic formation redshift, the general formalism does not
depend on the particular choice of $\PzM$ or the
corresponding interpretation of the quasar luminosity function.
A similar analysis may be repeated for other theoretical 
frameworks connecting galaxy
formation to supermassive black hole growth.

The largest final black holes masses we consider are
$\MBH \sim 3 \times 10^9 M_{\sun}$, which are produced
in the mergers of the very largest galaxies in our simulations.
In principle, the rarest gas-rich mergers at high-redshifts could produce
extraordinarily massive black holes.  The upper mass limit
for black holes in the simulations presented here only reflects the
final potential well depth of the largest galaxies we chose to merge as
the $\msigma$ relation generated by these mergers is a self-regulated 
process. The black holes grow in size to accommodate the growing potential
fomented by star formation in the highest density regions of the galaxy
during the height of the merger.  As the energy input from the black hole
into the surrounding gas saturates the ability of the gas contained by
the potential to cool, a strong outflow is driven from the central regions
and the growth of the black hole and spheroid are truncated.  There is no indication
in our simulations that this process cannot continue in suitably massive
gas-rich galaxy mergers to produce $\sim 10^{10} M_{\sun}$ black holes, and
future work to determine the possibility of forming such black holes in
galaxies mergers is planned.

However, evidence from the red sequence of the
galaxy color bimodality \citep{strateva01a,blanton03a,baldry04a,bell04a,faber05a} suggests that the
most massive galaxies are produced through the dry merging of spheroids at
$z<1$ \citep{bell05a,van_dokkum05a}.  In this picture, $10^{10} M_{\sun}$ black holes
would likely have to grow in mass partially through binary black hole mergers.  Whether black hole
coalescence can occur in collisionless systems on a reasonable timescale is
still an unresolved theoretical problem \cite[e.g.,][]{begelman80a,milosavljevic01a}.
Cosmological simulations that include prescriptions for black hole growth
may naturally account for the formation of the most massive black holes and
spheroids, but even this approach will likely experience limitations owing
to simulated volume size as the most massive galaxies are extremely rare.
Semi-analytical
models that include prescriptions for black hole feedback provide another promising alternative,
as such calculations can take advantage of N-body
simulations like the Millennium Run \citep{springel05d} that simulate cosmological volumes
with $10^{10}$ dark matter particles.  
However,
the semi-analytic approach may ultimately need to be calibrated against
simulations like those presented here if there is any hope of realistically
capturing the complicated hydrodynamical process that generates the $\msigma$
relation.  The existence and possible origin of the most massive SMBHs 
therefore promises to be an interesting avenue of future research.

Lastly, the robust nature of the $\msigma$ relation produced by our
simulations suggests that the merging of galaxies over cosmic time
will produce a relation with small intrinsic scatter, in accord
with observations.  The weak evolution of the calculated $\msigma$
normalization implies that the small observed scatter in the $\msigma$
reflects the robust physical origin of the relation
\citep[e.g.][]{haehnelt00a}.
While the dispersion $\Delta_{\MBH}=0.25-0.3$
measured by \cite{tremaine02a} is estimated to be 
intrinsic to the $\msigma$ relation, if the observational
errors have been significantly underestimated then our
physical model may need improvement.
We note that preliminary evidence of a
second parameter in the $\msigma$ relation \citep{siopis04a} may be
connected to statistical variance introduced by a redshift-dependent
normalization, which may take the form of either a correlation with
galaxy morphology or environment.
Further comparisons between the
observational data and simulations could help clarify these
possibilities, especially in a cosmological setting.  The redshift
evolution of the Faber-Jackson relation measured in the remnants may
bear on observations of the fundamental plane
\citep{dressler87a,djorgovski87a} and will be explored in forthcoming
work.

\section{Summary}
\label{section:summary}

We calculate the $\msigma$ relation produced by the
merging of disk galaxies appropriate for redshifts $z=0,2,3,$ and $6$
using a large set of $112$ hydrodynamic simulations that include the
effects of feedback from accreting supermassive black holes and
supernovae.
We develop a method for
comparing the $\msigma$ relation produced by simulations with existing
observational constraints through the statistical moments of the
measured relation using the formalism presented by \cite{yu04a},
and find the simulations to be in agreement with
the dispersion $\Delta_{\log \MBH}=0.25-0.3$ reported by
\cite{tremaine02a}.  The conclusions of our work are: 

\begin{itemize}

\item The simulations suggest that the normalization of the $\msigma$
relation may experience a weak evolution with redshift while the slope
is consistent with being constant near the locally observed value of
$\beta \sim 4$ \citep{ferrarese00a,gebhardt00a}.  As our simulations
cover a wide range of virial mass, gas fraction, ISM pressurization,
dark matter halo concentration, and surface mass density at each
redshift, our results suggest the physics that set the $\msigma$
scaling are remarkably insensitive to the properties of progenitor
galaxies in the scenario where the $\msigma$ relation is created
through coupled black hole and spheroid growth in galaxy mergers.

\item We demonstrate how to compare theoretical models for the
$\msigma$ relation that predict redshift evolution in either the slope
or normalization to the observed $\msigma$ relation by calculating the
$z=0$ dispersion $\Delta_{\log \MBH}$ of the relation given a functional
form for the evolution and a probability distribution function $\PzM$
that relates black hole mass to a characteristic formation redshift.
Adopting a model for $\PzM$ from \cite{hopkins05e}, we calculate the
allowed strength of redshift evolution given power-law, redshift evolution in
either normalization or slope consistent with the
observed scatter at $z=0$.
The modest normalization evolution
calculated by our simulations produces a scatter similar to that
observed.  We demonstrate how redshift evolution in the $\msigma$ relation
introduces mass-dependent dispersion and skewness about the average
relation, and show how these statistical moments can be used to
differentiate theoretical models for the $\msigma$ relation that
predict distinct redshift evolutions.

\item We relate our simulations to observations of the $\msigma$
relation in AGN at redshifts $z>0$ and, given the observational and
theoretical uncertainties, find the properties of our inactive
remnants to be more consistent with the redshift-independent $\msigma$
relation measured by \cite{shields03a} than the claims of
\cite{treu04a} and \cite{walter04a} that supermassive black
holes may form earlier than galactic spheroids.  We show that if the
offset $\delta \log \sigma=-0.16$ in the $\msigma$ relation measured
by \cite{treu04a} for AGN at $z=0.37$ is attributed to a power-law
evolution in the normalization of the velocity dispersion as
$\sigma(z) \propto \sigma(z=0)(1+z)^{\xi}$, the resultant scatter
expected in the $z=0$ $\msigma$ relation would be much larger than is
observed.  Additional comparisons between the active phase of our
merger simulations and the observations of active galaxies are planned
in future work.

\item Our simulations provide additional motivation for future observations of high-redshift normal galaxies using the next-generation of large telescopes for the purpose of constraining the evolution of the cosmic black hole population with time. The planned $20-30$m-class telescopes with infrared spectrographs should be able to measure central stellar velocity dispersions in galaxies to $z\sim3$. Our simulations can be used to connect those measurements to estimates for the supermassive black hole mass function and the distribution of black hole formation redshifts $\PzM$.
A complete census of the cosmic black hole population will require measurements of the spheroid population at high redshifts and will serve as important input into structure formation theories that account for the effects of black hole feedback on normal galaxies.
\end{itemize}

With a method for constraining theories of the $\msigma$ relation with the statistical moments of the local $\msigma$ relation based on their predictions for redshift evolution in the relation, future opportunities for direct comparisons between observations and theoretical models should be plentiful.  As the statistics of the observational results improve, the theoretical models will be able to adjust their physical prescriptions to better accommodate empirical constraints.  Research into the $\msigma$ relation should then continue to prove an important area of interest for future efforts to construct a more complete theory of galaxy formation.

\acknowledgements
We acknowledge the helpful comments of the referee, Doug Richstone, that helped
improve the presentation and clarity of our results.
This work was supported in part by NSF grants ACI 96-19019,  AST 00-71019,
AST 02-06299, and AST 03-07690,  and NASA ATP grants NAG5-12140, NAG5-13292, and
NAG5-13381.  The simulations were performed at the Center for Parallel Astrophysical Computing at Harvard-Smithsonian Center for Astrophysics.

\bibliographystyle{apj}

\end{document}